# Deriving the method of images for conductors

Chyi-Lung Lin, Hsien-Yi Lin

Department of Physics, Soochow University,

Taipei, Taiwan, 111, R.O.C.

We start from the simple fact that the method of images can always be used to obtain $\Phi_\sigma(\vec{r}_{in})$, which is the potential inside conductor produced by induced surface charges. We use this fact to construct image method for outside potential $\Phi_\sigma(\vec{r}_{out})$. We show that if we can find a relation between $\Phi_\sigma(\vec{r}_{out})$ and $\Phi_\sigma(\vec{r}_{in})$, then the image method for $\Phi_\sigma(\vec{r}_{in})$ can be used to derive the image method for $\Phi_\sigma(\vec{r}_{out})$. The number, the position and the amount of charge of the images can be directly derived. The discussion can be extended to the general n dimensions.

Chyi-Lung Lin:     cllin@scu.edu.tw

Hsien-Yi Lin:      hsienyi@scu.edu.tw

# INTRODUCTION

Solving the electrostatic potential, when the charge density is not fully known, in general, needs some special techniques, such as using separation of variables to solve the Poisson or Laplace equation, or multipole expansion, or the method of images [1-4].

The method of images for calculating potential outside a conductor has a long history. The typical one is a static charge q in front of a grounded spherical conductor. Without knowing the distribution of the induced charges on the surface, the potential outside can be calculated by q and an inside image charge. It seems this was first found by Sir William Thomson, later called Lord Kelvin in 1848 [5]. The method of images is, in fact, a very beautiful observation, except that it seems to lack of an intuitive derivation for this method. We try to fill this gap in this paper for the case of conductors.

For conductors in electrostatics, the method of images is to find a set of image charges to replace the complicate induced surface charges. We then concentrate on the potential due to surface charges. We let $\Phi_\sigma(\vec{r})$ represent the potential at $\vec{r}$ produced by the induced surface charges, and $\Phi_q(\vec{r})$ the potential by charge q. If q is located at $\vec{r}_q$, $\Phi_q(\vec{r})$ is given by

For n = 2 dimensions:  $\Phi_q(\vec{r}) = \frac{-q}{2\pi\epsilon_0} \text{Ln}[\mu |\vec{r} - \vec{r}_q|].$  (1)

For n = 3 dimensions:  $\Phi_q(\vec{r}) = \frac{q}{4\pi\epsilon_0} |\vec{r} - \vec{r}_q|^{-1}.$  (2)

We have put an arbitrary constant µ in (1), for the reason that the zero potential in two-dimension cannot be set at infinity.

Since our discussion can easily be extended to n > 3 dimensions, we also write the results for n ≥ 3. We have

For n ≥ 3 dimensions:  $\Phi_q(\vec{r}) = q\, \beta_n\, |\vec{r} - \vec{r}_q|^{2-n}$,    (3)

where $\beta_n = \frac{1}{\epsilon_0 n\, (n-2) c[n]}$ , n ≥ 3; $c[n] = \pi^{\frac{n}{2}}/\Gamma(\frac{n}{2}+1)$, and $\Gamma(n)$ is the Gamma function. We note that the volume of a sphere of radius a in n-dimension is $c[n]a^n$. For n = 2, $c[2] = \pi$. For n = 3, $c[3] = \frac{4\pi}{3}$, therefore $\beta_3 = \frac{1}{4\pi\epsilon_0}$. We define $\beta_2 = \frac{1}{2\pi\epsilon_0}$.

For $\Phi_\sigma(\vec{r})$, from the results of (1-3), we have:

For n = 2 dimensions:  $\Phi_\sigma(\vec{r}) = -\int dq_s\, \beta_2\, \mathrm{Ln}\,[\,\mu\,|\vec{r} - \vec{r}_s|\,]$,    (4)
For n ≥ 3 dimensions:  $\Phi_\sigma(\vec{r}) = \int dq_s\, \beta_n\, |\vec{r} - \vec{r}_s|^{2-n}$,    (5)

where the integration is over all the charges $dq_s$ distributed on the surface of conductor, and $\vec{r}_s$ is the position vector of $dq_s$. We also let $\vec{r}_{in}$ represent the position vectors of those points inside conductor, and $\vec{r}_{out}$ for points outside. Outside the conductor, the potential $\Phi_\sigma(\vec{r}_{out})$ is what we are trying to find out. The potential $\Phi_\sigma(\vec{r}_{in})$, in contrast to $\Phi_\sigma(\vec{r}_{out})$, is known and has simple solution. This is due to the simple fact that the method of images is always applicable for $\Phi_\sigma(\vec{r}_{in})$. To facilitate discussion, we consider an external point charge q which is located at $\vec{r}_q$ and is outside a grounded solid conductor, see Fig. 1. There is surface charge density $\sigma_s$ induced on the surface. We state below the simple fact of $\Phi_\sigma(\vec{r}_{in})$ in more detail.

**The simple fact for $\Phi_\sigma(\vec{r}_{in})$:**

Up to a constant, $\Phi_\sigma(\vec{r}_{in})$ can be obtained from the method of images. If the external point charge q is located at $\vec{r}_q$, then the image for calculating $\Phi_\sigma(\vec{r}_{in})$ is of charge (-q) and is just located at $\vec{r}_q$.

This can be proved as follows. Referring to Fig. 1, inside the conductor, the electric field at $\vec{r}_{in}$ is $\vec{E} = \vec{E}_\sigma(\vec{r}_{in}) + \vec{E}_q(\vec{r}_{in}) = 0$, where $\vec{E}_\sigma(\vec{r}_{in})$ is the electric field produced by induced surface charges, and $\vec{E}_q(\vec{r}_{in})$ is that by external charge q. Then $\vec{E}_\sigma(\vec{r}_{in}) = - \vec{E}_q(\vec{r}_{in})$. We see that $\vec{E}_\sigma(\vec{r}_{in})$ all point towards q, as if they are from a point charge (-q) located at $\vec{r}_q$. Thus the image method works for $\vec{E}_\sigma(\vec{r}_{in})$. Since $\vec{E} = 0$ inside the conductor, we have $\Phi_\sigma(\vec{r}_{in}) + \Phi_q(\vec{r}_{in})$ = constant. We define this constant by $\phi_0$, then $\Phi_\sigma(\vec{r}_{in}) = -\Phi_q(\vec{r}_{in}) + \phi_0 = \Phi_{-q}(\vec{r}_{in}) + \phi_0$. Thus, up to a constant, the potential $\Phi_\sigma(\vec{r}_{in})$ is from an image which is of charge (-q) and is located at $\vec{r}_q$. This proves the simple fact.

This simple fact for $\Phi_\sigma(\vec{r}_{in})$ may seem trivial, however, we will use it to construct the image method for $\Phi_\sigma(\vec{r}_{out})$. To simplify the discussions below, we neglect the constant $\phi_0$, which corresponds to the case of a grounded conductor. From the simple fact, we have the following results for $\Phi_\sigma(\vec{r}_{in})$ when the external charge q is located at $\vec{r}_q$:

For n = 2 dimensions:
$$\Phi_\sigma(\vec{r}_{in}) = -\Phi_q(\vec{r}_{in}) = q\, \beta_2\, \mathrm{Ln}[\, \mu\, |\vec{r}_{in} - \vec{r}_q\,|\,]. \qquad (6)$$

For n $\geq$ 3 dimensions:
$$\Phi_\sigma(\vec{r}_{in}) = -\Phi_q(\vec{r}_{in}) = -q\, \beta_n\, |\vec{r}_{in} - \vec{r}_q|^{2-n}. \qquad (7)$$

We may call formulas (6-7) the image-expression for $\Phi_\sigma(\vec{r}_{in})$. Formulas (6-7) are the key formulas for calculating $\Phi_\sigma(\vec{r}_{out})$. To calculate $\Phi_\sigma(\vec{r}_{out})$, our idea is that we try to relate $\Phi_\sigma(\vec{r}_{out})$ to $\Phi_\sigma(\vec{r}_{in})$. If this can be done, that means we have established a relation between $\vec{r}_{in}$ and $\vec{r}_{out}$. We define an inversion operator $\widetilde{P}$ for this connection. $\widetilde{P}$ maps one to one a vector $\vec{r}_{out}$ to a vector $\vec{r}_{in}$. That is $\widetilde{P}\,\vec{r}_{in} = \vec{r}_{out}$, and vice versa. Since $\Phi_\sigma(\vec{r}_{in})$ has an image-expression, $\Phi_\sigma(\vec{r}_{in})$ is related to $|\vec{r}_{in} - \vec{r}_q|$. Using the inversion operator $\widetilde{P}$, we transform $|\vec{r}_{in} - \vec{r}_q|$ into $|\vec{r}_{out} - \widetilde{P}\,\vec{r}_q|$, and therefore we come to the desired result that $\Phi_\sigma(\vec{r}_{out})$ is resulted from an image charge at $\widetilde{P}\,\vec{r}_q$. This describes the way we derive the method of images for $\Phi_\sigma(\vec{r}_{out})$. We illustrate our method in the following examples.

**Example-1. Grounded planar conductor in two-dimension**

We have a charge q located at (0, d) and a conductor in the half space $y \leq 0$, see Fig. 2. There is induced surface density $\sigma_s$ on the surface of the conductor. To calculate the outside potential, $\Phi_\sigma(\vec{r}_B)$, where B is a point outside the conductor, we note that the potential from $\sigma_s$ is symmetrical about the x-axis. Hence

$$\Phi_\sigma(\vec{r}_B) = \Phi_\sigma(\vec{r}_A), \qquad (8)$$

where A is the mirror image of B. We have then related $\Phi_\sigma(\vec{r}_{in})$ and $\Phi_\sigma(\vec{r}_{out})$. In this case, $\vec{r}_{in}$ and $\vec{r}_{out}$ are related by an inversion operator $\widetilde{P}$ which maps a point (x, y) to its mirror image, that is, $\widetilde{P}(x, y) = (x, -y)$. Then $\widetilde{P}\,\vec{r}_B = \vec{r}_A$, and $\widetilde{P}\,\vec{r}_A = \vec{r}_B$. We have the following properties of $\widetilde{P}$:

$$\widetilde{P}^2(x, y) = \widetilde{P}\,\widetilde{P}(x, y) = (x, y), \qquad (9)$$

$$|\vec{r}_1 - \vec{r}_2| = |\widetilde{P}\vec{r}_1 - \widetilde{P}\vec{r}_2|, \qquad (10)$$

where $\vec{r}_1$ and $\vec{r}_2$ are any two arbitrary vectors.

Since A is inside the conductor, $\Phi_\sigma(\vec{r}_A)$ is given by (6), we have

$$\Phi_\sigma(\vec{r}_B) = q\,\beta_2\,\mathrm{Ln}\,[\,\mu\,|\vec{r}_A - \vec{r}_q\,|\,]. \qquad (11)$$

We need to express $|\vec{r}_A - \vec{r}_q|$ in terms of $\vec{r}_B$. Using (10), $|\vec{r}_A - \vec{r}_q|$ $= |\widetilde{P}\vec{r}_A - \widetilde{P}\vec{r}_q| = |\vec{r}_B - \widetilde{P}\vec{r}_q|$. Formula (11) can then be written as

$$\Phi_\sigma(\vec{r}_B) = q\,\beta_2\,\mathrm{Ln}\,[\,\mu\,|\vec{r}_B - \widetilde{P}\vec{r}_q\,|\,]. \qquad (12)$$

Formula (12) states that the potential $\Phi_\sigma(\vec{r}_B)$ is produced by an image of charge (-q) located at $\widetilde{P}\vec{r}_q = (0,-d)$.

This example may seem trivial from the point of view of geometry symmetry. But the concept of inversion operator is important. It is this operator that takes us from $\vec{r}_q$ to $\widetilde{P}\vec{r}_q$, which is the location of the image charge.

**Example-2. Two semi-infinite grounded conducting surfaces meet at right-angles**

Referring to Fig.3, we have a conductor in the second, third and fourth quadrants, and we have a charge q located at the position $\vec{r}_q = (x_q, y_q)$, where $x_q > 0$ and $y_q > 0$. There are surface charges induced on the two meeting surfaces. We denote these two surfaces, respectively, by S1 and S2. And we define two inversion operators: $\widetilde{P}_1$ and $\widetilde{P}_2$, such that $\widetilde{P}_1$ (x,y) = (x,- y) and $\widetilde{P}_2$ (x,y) = (-x, y). We have $\widetilde{P}_1 \widetilde{P}_2$ (x, y) = (-x, - y) = $\widetilde{P}_2 \widetilde{P}_1$ (x, y). Both $\widetilde{P}_1$ and $\widetilde{P}_2$ satisfy the relations of (9-10). For a short notation, the potential $\Phi_\sigma(\vec{r})$ from a surface S is denoted by $\Phi_\sigma(\vec{r}|S)$. Since $\Phi_\sigma(\vec{r})$ comes from the contribution of S1 and S2, then $\Phi_\sigma(\vec{r}) = \Phi_\sigma(\vec{r}|S1 + S2)$.

As usual, let B be a point outside the conductor, then $\Phi_\sigma(\vec{r}_B) = \Phi_\sigma(\vec{r}_B|S1 + S2) = \Phi_\sigma(\vec{r}_B|S1) + \Phi_\sigma(\vec{r}_B|S2)$. We try to relate $\Phi_\sigma(\vec{r}_B)$ to an inside potential $\Phi_\sigma(\vec{r}_{in})$. Referring to Fig. 3, the first term $\Phi_\sigma(\vec{r}_B|S1)$, from the geometry symmetry, is equal to $\Phi_\sigma(\widetilde{P}_1\vec{r}_B|S1)$, and can be calculated as follows:

$$\Phi_\sigma(\vec{r}_B|S1) = \Phi_\sigma(\widetilde{P}_1\vec{r}_B|S1)$$
$$= \Phi_\sigma(\widetilde{P}_1\vec{r}_B|S1 + S2) - \Phi_\sigma(\widetilde{P}_1\vec{r}_B|S2)$$
$$= \Phi_\sigma(\widetilde{P}_1\vec{r}_B) - \Phi_\sigma(\widetilde{P}_2\widetilde{P}_1\vec{r}_B|S2), \quad (13)$$

where we use the fact that $\Phi_\sigma(\widetilde{P}_1\vec{r}_B|S2) = \Phi_\sigma(\widetilde{P}_2\widetilde{P}_1\vec{r}_B|S2)$, which is obvious from geometry symmetry. The same for the second term:

$$\Phi_\sigma(\vec{r}_B|S2) = \Phi_\sigma(\widetilde{P}_2\vec{r}_B) - \Phi_\sigma(\widetilde{P}_1\widetilde{P}_2\vec{r}_B|S1). \quad (14)$$

Adding (13) and (14), we have:

$$\Phi_\sigma(\vec{r}_B) = \Phi_\sigma(\widetilde{P}_1\vec{r}_B) + \Phi_\sigma(\widetilde{P}_2\vec{r}_B) - \Phi_\sigma(\widetilde{P}_1\widetilde{P}_2\vec{r}_B). \quad (15)$$

As $\widetilde{P}_1\vec{r}_B$, $\widetilde{P}_2\vec{r}_B$ and $\widetilde{P}_1\widetilde{P}_2\vec{r}_B$, are points inside the conductor, we have then related the outside potential $\Phi_\sigma(\vec{r}_B)$ to three inside potentials. We can then apply the image-expression for these three potentials. For example, for the first term, we have:

$$\Phi_\sigma(\widetilde{P}_1\vec{r}_B) = q\,\beta_2\,\text{Ln}\left[\,\mu\,\left|\widetilde{P}_1\vec{r}_B - \vec{r}_q\right|\,\right]. \tag{16}$$

Using the inversion operator $\widetilde{P}_1$, we have $\left|\widetilde{P}_1\vec{r}_B - \vec{r}_q\right| = \left|\vec{r}_B - \widetilde{P}_1\vec{r}_q\right|$, then (16) can be written as:

$$\Phi_\sigma(\widetilde{P}_1\vec{r}_B) = q\,\beta_2\,\text{Ln}\left[\,\mu\,\left|\vec{r}_B - \widetilde{P}_1\vec{r}_q\right|\,\right]. \tag{17}$$

The final result for (15) is therefore:

$$\Phi_\sigma(\vec{r}_B) = q\,\beta_2\,\text{Ln}\left[\mu\,|\,\vec{r}_B - \widetilde{P}_1\vec{r}_q\,|\right] + q\,\beta_2\,\text{Ln}\left[\mu\,|\,\vec{r}_B - \widetilde{P}_2\vec{r}_q\,|\right]$$
$$- q\,\beta_2\,\text{Ln}\left[\mu\,|\,\vec{r}_B - \widetilde{P}_1\widetilde{P}_2\vec{r}_q\,|\right]. \tag{18}$$

Formula (18) shows that $\Phi_\sigma(\vec{r}_B)$ can be obtained from three image charges, one is of charge (-q) located at $\widetilde{P}_1\vec{r}_q$, the other is of charge (-q) located at $\widetilde{P}_2\vec{r}_q$, and the third is of charge (+q) located at $\widetilde{P}_1\widetilde{P}_2\vec{r}_q$. Our calculation shows that $\Phi_\sigma(\vec{r}_B)$ is contained in $\Phi_\sigma(\widetilde{P}_1\vec{r}_B)$ and $\Phi_\sigma(\widetilde{P}_2\vec{r}_B)$. But the sum of these two potentials is over the value of $\Phi_\sigma(\vec{r}_B)$, and we need to subtract the amount of excess, which is the potential $\Phi_\sigma(\widetilde{P}_1\widetilde{P}_2\vec{r}_B)$. We note that the external charge is located at $\vec{r}_q$, and the image charges are located at $\widetilde{P}_1 \vec{r}_q$, $\widetilde{P}_2 \vec{r}_q$, and $\widetilde{P}_1 \widetilde{P}_2 \vec{r}_q$.

This method may also be applied in a similar way to obtain results for parallel plates or two half planes meeting at angles $\theta = \pi/\text{integer}$.

**Example-3. Grounded conducting sphere in two dimensions**

We consider a circular conductor of radius a and with center at the origin, see Fig. 4. An external charge q is at $\vec{r}_q = (x_q, 0)$, $x_q > a$. There is induced surface charge density $\sigma_s$. We consider the outside potential $\Phi_\sigma(\vec{r}_B)$, which is given by (4), with $\vec{r}_s = (a \cos[\theta_s], a \sin[\theta_s])$, where $\theta_s$ is the polar angle of $\vec{r}_s$, and $\vec{r}_B = (x_B, y_B)$ with $r_B = |\vec{r}_B| > a$. We have $r_s = |\vec{r}_s| = a$, and we denote an unit vector of $\vec{r}$ by $\hat{r}$.

We try to relate $\Phi_\sigma(\vec{r}_B)$ to $\Phi_\sigma(\vec{r}_{in})$, so that we can apply the result of (6). Thus, we need to express the term $|\vec{r}_B - \vec{r}_s|$ in (4) in terms of an inside vector $\vec{r}_{in}$. We use the trick: $|\vec{r}_B - \vec{r}_s| = |r_B \hat{r}_B - a \hat{r}_s| = |a \hat{r}_B - r_B \hat{r}_s| = \frac{r_B}{a} | \frac{a^2}{r_B^2} \vec{r}_B - \vec{r}_s |$. We then define

$$\vec{r}_A = \frac{a^2}{r_B^2} \vec{r}_B. \qquad (19)$$

$\vec{r}_A$ represents the position vector of a point A. The point A is on the line $\overrightarrow{OB}$ and is inside the conductor, as $r_A = \frac{a^2}{r_B} < a$. We have then the interesting formula for all surface vectors $\vec{r}_s$:

$$|\vec{r}_B - \vec{r}_s| = \frac{r_B}{a} | \vec{r}_A - \vec{r}_s |. \qquad (20)$$

From (19) we define the inversion operator $\widetilde{P}$ of a circle as:

$$\widetilde{P} \vec{r} = \frac{a^2}{r^2} \vec{r}. \qquad (21)$$

We see that $\widetilde{P}$ maps a point outside a circle to a point inside, and vice versa. Thus $\widetilde{P} \vec{r}_B = \vec{r}_A$ and $\widetilde{P} \vec{r}_A = \vec{r}_B$. Also we have:

$$\widetilde{P}^2 \vec{r} = \vec{r},$$
$$\widetilde{P} \vec{r}_s = \vec{r}_s. \qquad (22)$$

There is also the following important formula for two arbitrary vectors $\vec{r}_1$ and $\vec{r}_2$:

$$|\vec{r}_1 - \vec{r}_2| = \frac{r_1 r_2}{a^2} \; |\widetilde{P} \vec{r}_1 - \widetilde{P} \vec{r}_2|. \qquad (23)$$

We note that the relations (19-23) are independent of the dimensions. Hence the relations (19-23) also hold for dimensions $n \geq 3$. We will use these relations in the next example. Using (20), we can then calculate $\Phi_\sigma(\vec{r}_B)$ from (4) in the following way:

$$\Phi_\sigma(\vec{r}_B) = -\int \beta_2 \, dq_s \, \text{Ln} \, [\, \mu \, |\vec{r}_B - \vec{r}_s \,| \,]$$
$$= -\int \beta_2 \, dq_s \, \text{Ln} \, [\, \frac{r_B}{a} \, \mu \, |\, \vec{r}_A - \vec{r}_s \,| \,]$$
$$= -\int \beta_2 \, dq_s \, \text{Ln} \, \left[\frac{r_B}{a}\right] - \int \beta_2 \, dq_s \, \text{Ln} \, [\, \mu \, |\, \vec{r}_A - \vec{r}_s \,|\,]$$
$$= -\int \beta_2 \, dq_s \, \text{Ln} \, \left[\frac{r_B}{a}\right] + \Phi_\sigma(\vec{r}_A). \qquad (24)$$

We then relate the outside potential at $\vec{r}_B$ to the inside potential at $\vec{r}_A$, where $\vec{r}_A = \widetilde{P} \vec{r}_B$. From (24), letting $Q_s = \int dq_s$ = the total charge on the surface, and applying the image-expression for $\Phi_\sigma(\vec{r}_A)$, we have

$$\Phi_\sigma(\vec{r}_B) = -Q_s\,\beta_2\,\mathrm{Ln}\left[\frac{r_B}{a}\right] + q\,\beta_2\,\mathrm{Ln}\,[\mu\,|\,\vec{r}_A - \vec{r}_q\,|]. \qquad (25)$$

From (23), $|\vec{r}_A - \vec{r}_q| = \frac{r_A\,x_q}{a^2}|\vec{r}_B - \widetilde{P}\vec{r}_q| = \frac{x_q}{r_B}|\vec{r}_B - \widetilde{P}\vec{r}_q|$, then

$$\Phi_\sigma(\vec{r}_B) = -Q_s\,\beta_2\,\mathrm{Ln}\left[\frac{r_B}{a}\right] + q\,\beta_2\mathrm{Ln}\,[\frac{x_q}{r_B}\mu\,|\,\vec{r}_B - \widetilde{P}\vec{r}_q\,|]$$

$$= -Q_s\,\beta_2\,\mathrm{Ln}\left[\frac{r_B}{a}\right] + q\,\beta_2\,\mathrm{Ln}\left[\frac{x_q}{r_B}\right] + q\,\beta_2\,\mathrm{Ln}\,[\mu\,|\,\vec{r}_B - \widetilde{P}\vec{r}_q\,|]. \quad (26)$$

The third term of the potential is from an image of charge (-q) located at the position $\widetilde{P}\vec{r}_q = (\frac{a^2}{x_q}, 0)$. The first term of the potential contains the total charge $Q_s$. Since the charge of the image is (-q), we set $Q_s = -q$. Then we have:

$$\Phi_\sigma(\vec{r}_B) = q\,\beta_2\,\mathrm{Ln}\left[\frac{x_q}{a}\right] + q\,\beta_2\,\mathrm{Ln}[\mu\,|\,\vec{r}_B - \widetilde{P}\vec{r}_q\,|]. \qquad (27)$$

Thus, up to a constant, $\Phi_\sigma(\vec{r}_B)$ is from an image of charge (-q) located at $\widetilde{P}\vec{r}_q$. The constant $q\,\beta_2\,\mathrm{Ln}\left[\frac{x_q}{a}\right]$ is in fact needed. Since we set the potential being zero on the surface of the conductor, we should have

$$\Phi_\sigma(\vec{r}_s) + \Phi_q(\vec{r}_s) = 0. \qquad (28)$$

This is indeed so. As we note that

$$\Phi_q(\vec{r}_s) = -q\,\beta_2\,\mathrm{Ln}[\mu\,|\,\vec{r}_s - \vec{r}_q\,|].$$

$$\Phi_\sigma(\vec{r}_s) = q\,\beta_2\,\mathrm{Ln}\left[\frac{x_q}{a}\right] + q\,\beta_2\,\mathrm{Ln}[\mu\,|\,\vec{r}_s - \widetilde{P}\vec{r}_q\,|]. \quad (29)$$

Using (20) with $\vec{r}_B$ replaced by $\vec{r}_q$, and $\vec{r}_A$ replaced by $\widetilde{P}\vec{r}_q$, we have

$$|\vec{r}_s - \vec{r}_q| = \frac{x_q}{a}|\vec{r}_s - \widetilde{P}\vec{r}_q|. \qquad (30)$$

Taking this relation to (29), we easily check that (28) is satisfied. In conclusion, the induced surface charges provide a potential that is from an image of charge (-q) located at $\widetilde{P}\vec{r}_q$, and also provide a constant potential. This constant potential is to ensure the zero potential on the surface.

We note that if the conductor is of the shape as an ellipse, then the surface point with position vector $\vec{r}_s$ is of a magnitude $r_s$ which now is not a constant. The trick we use before now gives $|\vec{r}_B - \vec{r}_s| = \frac{r_B}{r_s}|\vec{r}_A - \vec{r}_s|$, where $\vec{r}_A = \frac{r_s^2}{r_B^2}\vec{r}_B$. We note that $\vec{r}_A$ is not a fixed vector, it varies with $\vec{r}_s$. Therefore we do not obtain a unique inversion point of B. We cannot find an inversion operator $\widetilde{P}$ connecting one to one outside points and inside points. The method of images constructed in this way then fails.

**Example-4. Grounded conducting sphere in N ≥ 3 dimensions**

We consider a spherical conductor of radius a in n dimensions and with center at the origin. An external charge q is located at $\vec{r}_q = (x_q, 0, 0, \ldots, x_n = 0)$, with $x_q > a$, see Fig. 4. The calculation of $\Phi_\sigma(\vec{r}_B)$ in this case is even simpler. We note that the relations (19-23) also hold for n ≥ 3 dimensions. Using these results, the outside potential $\Phi_\sigma$ at $\vec{r}_B$ is:

$$\Phi_\sigma(\vec{r}_B) = \int \beta_n \, dq_s \, |\vec{r}_B - \vec{r}_s|^{2-n}$$

$$= \int \beta_n \, dq_s \, (\tfrac{r_B}{a})^{2-n} |\vec{r}_A - \vec{r}_s|^{2-n}$$

$$= (\tfrac{r_B}{a})^{2-n} \Phi_\sigma(\vec{r}_A), \qquad (31)$$

where $\vec{r}_A$ is defined in (19). Thus we have related $\Phi_\sigma(\vec{r}_{out})$ to $\Phi_\sigma(\vec{r}_{in})$. Note that the proportional constant $(\tfrac{r_B}{a})^{2-n}$ depends on $r_B$. Using (7) and (23), we have

$$\Phi_\sigma(\vec{r}_B) = -(\tfrac{r_B}{a})^{2-n} q \, \beta_n \, |\vec{r}_A - \vec{r}_q|^{2-n}$$

$$= -(\tfrac{r_B}{a})^{2-n} q \, \beta_n \, (\tfrac{r_A \, x_q}{a^2})^{2-n} |\vec{r}_B - \widetilde{P}\vec{r}_q|^{2-n}$$

$$= -\left(\tfrac{a}{x_q}\right)^{n-2} q \, \beta_n \, |\vec{r}_B - \widetilde{P}\vec{r}_q|^{-(n-2)}. \qquad (32)$$

We see that $\Phi_\sigma(\vec{r}_B)$ is from an image of charge $-\left(\tfrac{a}{x_q}\right)^{n-2} q$, located at $\widetilde{P}\vec{r}_q = (\tfrac{a^2}{x_q}, 0, 0, \ldots, x_n = 0)$. For n = 3, the image is of charge $(-\tfrac{a}{x_q} q)$ and is located at $(\tfrac{a^2}{x_q}, 0, 0)$; this is well-known.

We easily check the zero potential on the conductor. We have

$$\Phi_q(\vec{r}_s) = q \, \beta_n \, |\vec{r}_s - \vec{r}_q|^{-(n-2)}$$

$$\Phi_\sigma(\vec{r}_s) = -\left(\tfrac{a}{x_q}\right)^{n-2} q \, \beta_n \, |\vec{r}_s - \widetilde{P}\vec{r}_q|^{-(n-2)}. \qquad (33)$$

Using (30), as that also holds for n ≥ 3 dimensions, we easily have that $\Phi_\sigma(\vec{r}_s) + \Phi_q(\vec{r}_s) = 0$.

**Conclusions**

To calculate $\Phi_\sigma(\vec{r}_{out})$ by the method of images, our point is that we start from the simple fact that $\Phi_\sigma(\vec{r}_{in})$ can be obtained from the method of images. We then try to relate $\Phi_\sigma(\vec{r}_{out})$ to $\Phi_\sigma(\vec{r}_{in})$. If this can be done, that means we have found an inversion operator $\widetilde{P}$ which relates $\vec{r}_{in}$ and $\vec{r}_{out}$. Then the image method for $\Phi_\sigma(\vec{r}_{in})$ can be extended to derive the image method for $\Phi_\sigma(\vec{r}_{out})$. If $\vec{r}_q$ is the location of an external charge, it is also the location of the image charge for $\Phi_\sigma(\vec{r}_{in})$. Our result is that $\widetilde{P}\vec{r}_q$ is the location of the image charge for $\Phi_\sigma(\vec{r}_{out})$. For more complex cases, $\Phi_\sigma(\vec{r}_{out})$ may be related to a set of inside potentials. It then needs a set of inversion operators $\{\widetilde{P}_i | i = 1,2,...N\}$, and therefore producing a set of image charges for $\Phi_\sigma(\vec{r}_{out})$. The locations of these images are expressed in terms of $\widetilde{P}_i$ and their combinations. Our method does not need any trial and error in determining the number, the position and the amount of charge of the images.

**Figure-1**

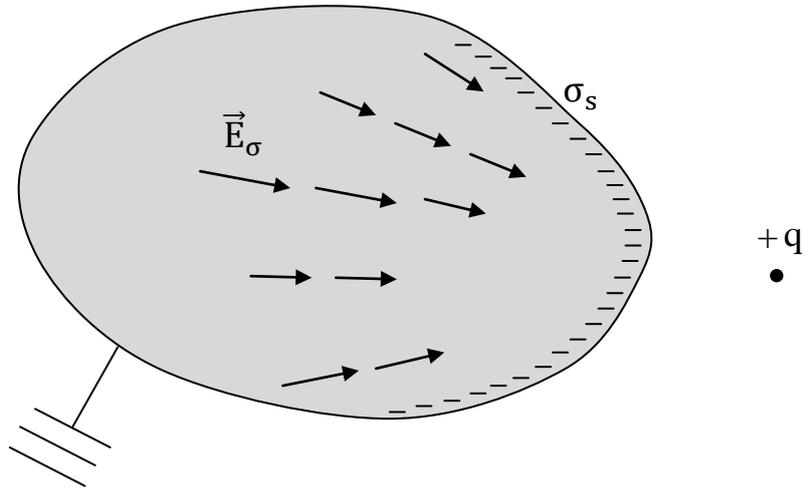

**Figure-2**

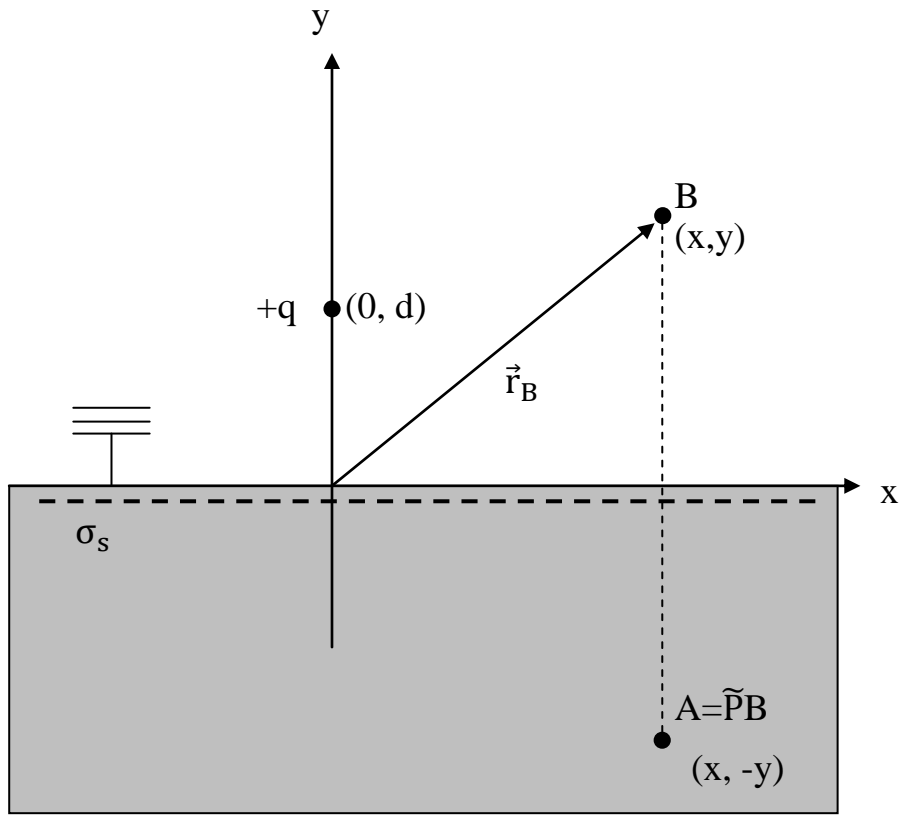

**Figure-3**

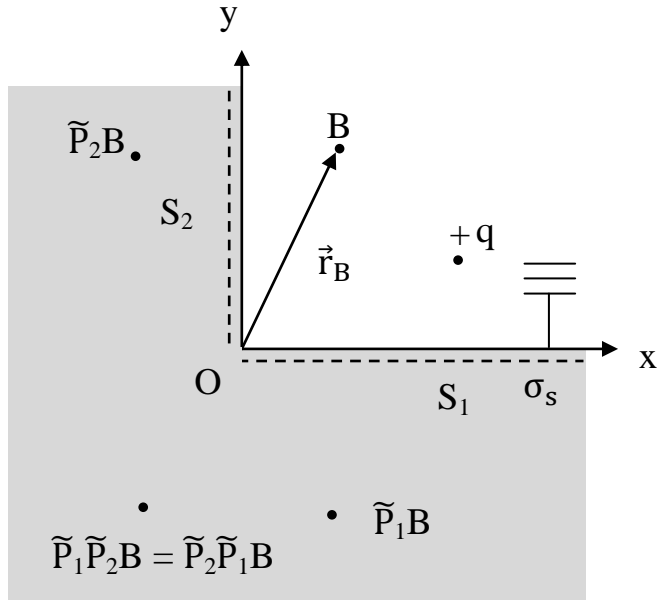

**Figure-4**

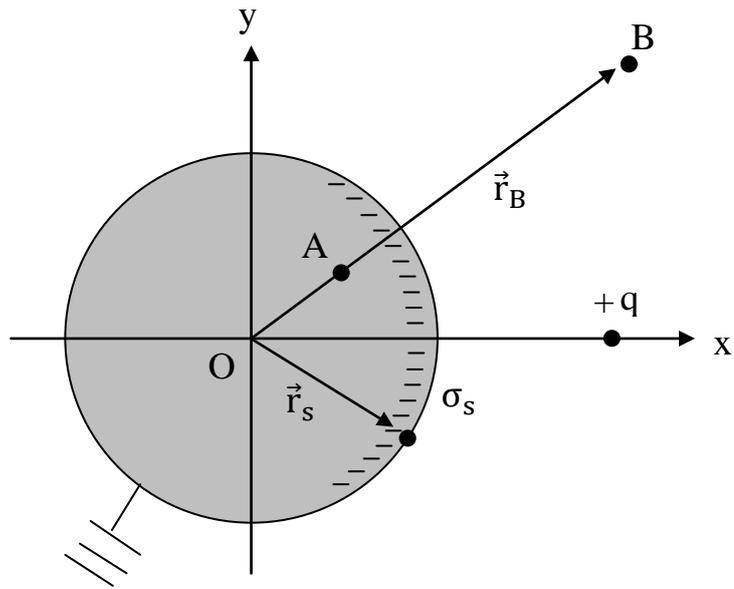

# FIGURE CAPTIONS

**Fig. 1:**

an external charge q is in front of a conductor of an arbitrary shape. The induced surface charge density $\sigma_s$ generates electric field $\vec{E}_\sigma$ inside the conductor. $\vec{E}_\sigma$ all point towards q, as if they are from a point charge (-q) located at the same position of the external charge q. Thus inside a conductor, the induced surface charges can always be replaced by an outside image charge for calculating the potential and electric field.

**Fig. 2:**

A charge q is located at (0,d). Below the x-axis is the conductor. Point B is a point outside the conductor, and A is the mirror image of B.

**Fig. 3:**

A charge q is located at the first quadrant. The conductor is in the other three quadrants.

**Fig. 4:**

an external charge q is in front of a spherical conductor of radius a with center at the origin. The point A is related to B by the inversion operator $\widetilde{P}$ defined in (21).